# On plasma density blobs in drift turbulence

S. I. Krasheninnikov

University California San Diego, 9500 Gilman Dr., La Jolla, California, 92093-0411

**Abstract**

By keeping nonlinear Boltzmann factor in electron density dependence on electrostatic potential it is demonstrated that large plasma density blobs, often seen in experiment inside separatrix, can exist within the framework of drift wave dynamics. The estimates show that plasma density in a blob can be ~3 times higher that average plasma density, but hardly exceeds this limit, which in a ball park is in agreement with experimental observations.

It is known that intermittent blobby transport plays a very important role in the transport of edge plasma in magnetic fusion devices (e.g. see Ref. 1-3). Although the mechanism, propelling high plasma density filaments (blobs) on the outer side of a tokamak in the Scrape-Off-Layer (SOL) is rather well understood [1-4], the physics responsible for the formation of the blobs is still under the questions. Meanwhile there is compelling experimental evidence that high-density blobs exist already inside the separatrix (e.g. see Ref. 5-7) where they move mainly in poloidal direction and once in a while cross the separatrix and appear in the SOL. In what follows we demonstrate that it is plausible that the formation of these high plasma density blobs is inherent for the drift wave plasma turbulence. To do this we start with the reconsideration of the Cherny-Hasegwa-Mima equation [8] and will not pursue a standard assumption about smallness of electrostatic fluctuations.

Following [9] we consider Boltzmann electrons

$$n_e(\vec{r},t) = \hat{n}(\vec{r}_\perp)\exp\{\phi(\vec{r},t)\}, \qquad (1)$$

(where $\phi = e\varphi/T_e$, e is the electron charge, $\varphi$ is the electrostatic potential, $T_e$ is the electron temperature which we assumed to be constant, and $\hat{n}(\vec{r}_\perp)$ describes electron density in the absence of electrostatic potential) and ion continuity equation

$$\frac{\partial n_i}{\partial t} + \nabla \cdot (n_i \vec{V}_i) = 0, \qquad (2)$$

where we assume cold ion approximation for ion velocity

$$\vec{V}_i = \vec{V}_0 - \rho_s^2 \frac{d}{dt}\nabla\phi \quad \text{and} \quad \vec{V}_0 = -D_B\left(\nabla\phi \times \vec{b}\right), \quad \frac{d(...)}{dt} = \frac{\partial(...)}{\partial t} + (\vec{V}_0 \cdot \nabla)(...), \qquad (3)$$

where $\rho_s^2 = T_e M(c/eB)^2$, $D_B = cT_e/eB$, M is the ion mass and B is the strength of the magnetic field. After we take $\hat{n}(\vec{r}_\perp) = n_0 \exp(-\Lambda x)$ and, using the quasi-neutrality condition, substitute the expressions (1) and (3) into Eq. (2) we find

$$\frac{de^\phi}{dt} - \rho_s^2 \nabla \cdot \left(e^\phi \frac{d}{dt}\nabla\phi\right) - \Lambda e^\phi \vec{e}_x \cdot \vec{V}_0 = 0, \qquad (4)$$

which we'll call drift-wave with non-linear electrons (NLE) equation. Assuming that $|\phi| \ll 1$ from Eq. (4) we obtain the Hasegawa-Mima equation (HM):

$$\frac{d}{dt}\left(\phi - \rho_s^2 \nabla^2 \phi\right) - \Lambda \vec{e}_x \cdot \vec{V}_0 = 0. \qquad (5)$$

Let us consider these equations for the case where $|\partial(...)/\partial y| \gg |\partial(...)/\partial x|$. For NLE we find

$$\frac{\partial e^\phi}{\partial t} - \rho_s^2 \frac{\partial}{\partial y}\left(e^\phi \frac{\partial^2 \phi}{\partial y \partial t}\right) + \Lambda D_B \frac{\partial e^\phi}{\partial y} = 0, \tag{6}$$

while for HM

$$\frac{\partial}{\partial t}\left(\phi - \rho_s^2 \frac{\partial^2 \phi}{\partial y^2}\right) + \Lambda D_B \frac{\partial \phi}{\partial y} = 0. \tag{7}$$

From Eq. (6,7) we see that the NLE conserves $\langle \exp(\phi)\rangle_y = \text{const.}$ while the HM conserves $\langle \phi \rangle_y = \text{const.}$, here $\langle ... \rangle_y$ is just averaging over coordinate y. This implies that both NLE and HM maintain a constant averaged density along y. We also notice that Eq. (7) following from the HM model is linear, while Eq. (6) obtained from NLE is still nonlinear.

Let us consider traveling wave, $\partial(...)/\partial t = -U_{HM/NLE}\partial(...)/\partial y$, solutions of Eq. (6,7) satisfying conditions $\langle \phi \rangle_y = 0$ for HM and it's analog $\langle \exp(\phi)\rangle_y = 1$ for NLE.

For HM we find just simple sinusoidal oscillation of $\phi$ with amplitude $\phi_0$, which formally should be small

$$\phi = \phi_0 \sin(\kappa_{HM} y), \tag{8}$$

where $\kappa$ is the effective wave number satisfying the following inequality

$$\kappa_{HM}^2 = \rho_s^{-2}(\Lambda D_B / U_{HM} - 1) > 0, \tag{9}$$

which requires $U_{HM} < \Lambda D_B$.

For traveling wave solution for the NLE equation we arrive to the following eqution

$$\kappa_{NLE}^2 \frac{de^\phi}{dy} + \frac{d}{dy}\left(e^\phi \frac{d^2 \phi}{dy^2}\right) = 0, \tag{10}$$

where

$$\kappa_{NLE}^2 = \rho_s^{-2}(\Lambda D_B / U_{NLE} - 1), \tag{11}$$

which at this moment may be both positive and negative. From Eq. (10) we find

$$\frac{d^2 \phi}{dy^2} = -\kappa_{NLE}^2 + Ce^{-\phi} \equiv -\frac{\partial}{\partial \phi} W(\phi), \tag{12}$$

where C is the integration constant and

$$W(\phi) = \kappa_{NLE}^2 \phi + C(e^{-\phi} - 1) ,\tag{13}$$

is the effective potential. From Eq. (12) we find the first integral

$$\frac{1}{2}\left(\frac{d\phi}{dy}\right)^2 + W(\phi) = E ,\tag{14}$$

where E is the effective energy. Therefore, as always, it is convenient to analyse Eq. (12), as the equation of motion of quasi-particle considering $\phi$ it's the effective coordinate and y as the effective time.

Since we have to satisfy $\langle \exp(\phi) \rangle_y = 1$, our solution, which we assume to be periodic, should be bounded by some "turning points" (where $d\phi/dy = 0$) at positive $\phi_{max}$ and at negative $\phi_{min}$ (if both $\phi_{max}$ and $\phi_{min}$ have the same sign $\langle \exp(\phi) \rangle_y = 1$ cannot be held) satisfying the conditions

$$W(\phi_{max}) = W(\phi_{min}) = E .\tag{15}$$

Moreover, since the existence of the solution (14) requires $W(\phi_{min} < \phi < \phi_{max}) < E$, ensuring that our quasi-particle moves across the "potential well", from Eq. (13) one sees that it is only possible when both $\kappa_{NLE}^2$ and C are positive, which implies $U_{NLE} < \Lambda D_B$. Finally, using Eq. (14), it is easy to show that we can express the condition $\langle \exp(\phi) \rangle_y = 1$ only in terms of $\phi$. For this purpose it is convenient to introduce $\hat{E} = E/\kappa_{NLE}^2$, $\hat{C} = C/\kappa_{NLE}^2$, and

$$\hat{W}(\phi) = \phi + \hat{C}(e^{-\phi} - 1) ,\tag{16}$$

as a result we arrive to the following equation

$$\int_{\phi_{min}}^{\phi_{max}} \frac{e^\phi d\phi}{\sqrt{\hat{E} - \hat{W}(\phi)}} = \int_{\phi_{min}}^{\phi_{max}} \frac{d\phi}{\sqrt{\hat{E} - \hat{W}(\phi)}} .\tag{17}$$

Thus, to find the solution of Eq. (12) which corresponds to our constrain $\langle \exp(\phi) \rangle_y = 1$, we need to find the solution of Eq. (17) and then use the "turning" points $\phi_{max}$ (or $\phi_{min}$) as the boundary conditions in Eq. (12). We will consider $\hat{C}$ as the input parameter and will try to find $\hat{E}(\hat{C})$ as the solution of Eq. (17).

To get some insights let we start with analysis of the potential $\hat{W}(\phi)$. First we note that is has only one extremum (minimum) at $\phi = \phi_{ext} = \ln(\hat{C})$. Depending on the magnitude of $\hat{C}$, $\phi_{ext}$ can be both positive ($\hat{C} > 1$) and negative ($\hat{C} < 1$), but $\hat{W}(\phi_{ext}) = \ln(\hat{C}) - \hat{C} + 1$ is always negative except the case $\hat{C} = 1$, where $\hat{W}(\phi_{ext}) = 0$. Schematically $\hat{W}(\phi)$ for different $\hat{C}$ is shown in Fig.1.

From Fig. 1 one can see that for $\hat{C} \gg 1$ quasi-particle spends most of the time at large $\phi > 0$. Therefore, taking into account that in this case $\exp(\phi) \gg 1$, we conclude the solution of Eq. (17) exist only for $\hat{C} < \hat{C}_{crit}$. The magnitude of $\hat{C}_{crit}$ can be found numerically, but this goes beyond the scope of this paper.

For $\hat{C} = 1$ and small $\phi$ we have $\hat{W}(\phi) = \phi^2 / 2$. Expanding $\exp(\phi) \approx 1 + \phi$ in the left hand side of Eq. (17) we find the equality holds for any $\hat{E}$. Of cause, in practice, taking into account both "anharmonic" features of $\hat{W}(\phi)$ and only approximate validity of expansion $\exp(\phi) \approx 1 + \phi$, some "quantization" of $\hat{E}$ will appear which will limit the maximum value of $\hat{E}$. However, it is very likely that for $\hat{C} \approx 1$ multiple solutions exist and they correspond to the "continuum" of the solutions described by Eq. (8) for the case of HM.

For $\hat{C} \ll 1$ $\hat{W}(\phi)$ is strongly squeezed toward negative $\phi$ where quasi-particle spends most of it's time. As a result, due to presence of $\exp(\phi)$ in the left hand side of Eq. (18), the regions of major contributions to the left and right hand sides of Eq. (17) are separated. While the major contribution to the left hand side gives the region of positive $\phi$, the magnitude of the integral on the right hand side is mainly determined by negative $\phi$. Then for $\hat{C} \ll 1$ we find $\phi_{max} \approx \hat{E}$ and

$$\int_{\phi_{min}}^{\phi_{max}} \frac{e^{\phi} d\phi}{\sqrt{\hat{E} - \hat{W}(\phi)}} \approx \sqrt{\pi} \exp(\hat{E}). \tag{18}$$

To estimate the integral on the right hand side we take into account fast growth of exponent in the expression for $\hat{W}(\phi)$ we can take $\phi_{min} \approx \alpha \phi_{ext} = \alpha \ln(\hat{C})$, where $\alpha \gtrsim 1$ is some numerical factor. As a result we have

$$\int_{\phi_{min}}^{\phi_{max}} \frac{d\phi}{\sqrt{\hat{E} - \hat{W}(\phi)}} \approx 2\sqrt{\hat{E} - \alpha \ell n(\hat{C})} \ . \tag{19}$$

Then from the expressions (18, 19) we have the following approximate solution

$$\exp(\hat{E}) \approx 2\sqrt{-(\alpha/\pi)\ell n(\hat{C})} \ . \tag{20}$$

Recalling that $\phi_{max} \approx \hat{E}$ and $n_e \propto \exp(\phi)$ we conclude that this solution corresponds to a strong local increase of plasma density, density blob, in comparison to the average one, up to

$$n_{blob}/\langle n_e \rangle \approx 2\sqrt{-(\alpha/\pi)\ell n(\hat{C})} \ . \tag{21}$$

These plasma density blobs are surrounded by plasma with depleted density. Since $\phi_{min} \approx \ell n(\hat{C})$ we have virtually plasma hole, where plasma density, $n_{hole}$, drops up to

$$n_{hole}/\langle n_e \rangle \approx \hat{C} << 1 \ . \tag{22}$$

Using Eq. (14) and our estimate (19) we find the distance between two consecutive blobs, $L_2$,

$$L_2 \approx 2^{3/2}\sqrt{-\alpha \ell n(\hat{C})}\Big/\kappa_{NLE} \ . \tag{23}$$

Defining the effective size of the blob, $\delta_b$, as the distance where $\phi > 0$, from Eq. (14, 20) we find

$$\delta_b \approx \sqrt{2}\ell n\left(2\sqrt{-(\alpha/\pi)\ell n(\hat{C})}\right)\Big/\kappa_{NLE} \ . \tag{24}$$

Comparing the expressions (23, 24) we conclude that for $\hat{C} << 1$ $\delta_b << L_2$ so that the blob indeed looks like a solitary structure similar to experimental observations made with the Gas Puff Imaging (GPI) technique in Ref. 5, 6. In addition, we notice that the ratio $n_{blob}/\langle n_e \rangle$ from Eq. (21) has a very weak dependence on the parameter $\hat{C}$ and taking, as an example, $\hat{C} \sim 0.1$ we find $n_{blob}/\langle n_e \rangle \sim 3$, which, again, seems to agree with the GPI observations.

In conclusion, by keeping nonlinear Boltzmann factor in electron density dependence on electrostatic potential we demonstrated that large plasma density blobs, often seen in experiment inside separatrix, can exist within the framework of drift wave dynamics. Our estimates show that can be ~3 times higher that average plasma density, but hardly exceeds this limit, which in a ball park is in agreement with experimental observations. Of cause more work still needed to learn the dynamics of the formation of these structures.

**Acknowledgements.** The author thanks Prof. A. I. Smolyakov for fruitful discussions. This material is based upon the work supported by the U.S. Department of Energy, Office of Science, Office of Fusion Energy Sciences under Award No. DE-FG02-04ER54739 at UCSD

**Figure captures**

Fig. 1. Schematic view of the function $\hat{W}(\phi)$ for different $\hat{C}$.

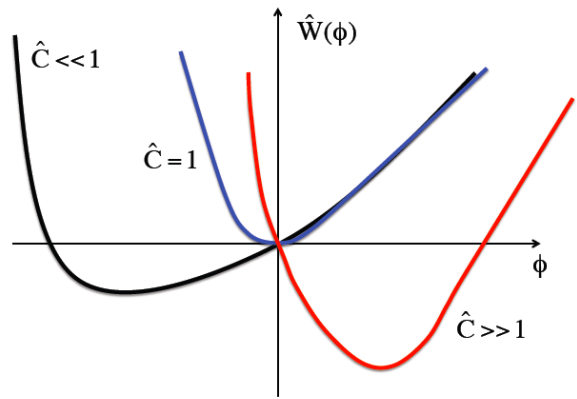

Fig. 1.